\documentstyle[pra,aps]{revtex}
\begin{document}
\draft

\twocolumn[\hsize\textwidth\columnwidth\hsize\csname 
@twocolumnfalse\endcsname

\title{Zeta Function Method for Repulsive Casimir Forces at Finite Temperature}

\author{F. C. Santos, A. Ten\'orio and A. C. Tort\cite{email}}

\address{Instituto de F\'{\i}sica, Universidade Federal do Rio de Janeiro,
Caixa Postal 68528, 21945-970 Rio de Janeiro RJ, Brazil}

\date{\today}
\maketitle

\begin{abstract}
We compute the Casimir energy between an unusual pair of parallel plates  at finite temperature, namely, a perfectely conducting plate ($\epsilon\rightarrow\infty$) and an infinitely permeable one ($\mu\rightarrow\infty$) by applying the generalized zeta function method. We also compute the Casimir pressure and discuss the high and the low temperature limits.
\end{abstract}

\vskip2pc] \label{sec:level1}

\section{Introduction}
Since Casimir's paper \cite{Ca} on the attraction between two parallel perfectly conducting plates due to the vacuum fluctuations of the electromagnetic field was published, a considerable amount of effort, which varies from the investigation of new geometries and theories to the application of the Casimir effect to alternative technologies, has been put into the study of this importante subject.  (For a review see, for example, Mostepanenko and Trunov \cite{MosteTrunov} or Plunien {\it et al}). Recently, the experimental observation of this effect was greatly improved by the experiment due to Lamoreaux \cite{Lamo}. From the theoretical viewpoint Casimir's approach to this problem essentially consisted in computing the interaction energy between the plates as the regularized difference between the zero point energies with and without boundary conditions dictated by the physical situation at hand, for instance, the perfectly conducting character of the plates. The great novelty!
 indeed of Casimir's 1948 paper was not the fact that two neutral object were atracted to each other, which was familiar to those studying dispersive van der Waals forces, but the simplicity of the method of calculating this attraction in the framework of qu
antum field theory.

 Casimir's definition of the vacuum energy requires a regularization recipe for its implementation. Many regularization techniques are available nowadays and, depending on the specific physical situation at hand, one of them may be more suitable than the others. Particularly, methods of computing effective actions are in general very powerful to give physical meaning to the divergent quantities we must deal with. Here we will be concerned with one of these methods, namely, the so-called generalized zeta function method \cite{ZETA}. We will apply it to the case of a pair of parallel infinite plates one of which is perfectly conducting ($\epsilon\to\infty$), while the other is infinitely permeable ($\mu\to\infty$). The setup will be considered to be in thermal equilibrium with a heat reservoir at finite temperature T. This problem at zero temperature was analyzed by Boyer \cite{Boyer} two decades ago in the framework of random electrodynamics, a kind of classical electrodynamics!
 whi
ch includes the zero-point electromagnetic radiation. Boyer was able to show that for this unusual pair of plates the Casimir energy is positive which results in a repulsive force per unit area between the plates. Here we shall study this problem within the framework of finite temperature QFT in order to take into account the thermal effects in equilibrium. We will employ units such that Boltzmann constant, the speed of light and Planck's constant are set equal to unity.
\section{Evaluation of the free energy}
Since we will be dealing with a system in thermal equilibrium, the imaginary time formalism will be convenient. In order to apply the generalized zeta function method, let us introduce the partition function ${\cal Z}$ for a bosonic theory \cite{Kapusta}:
\begin{equation}
{\cal Z} = N\int_{\mbox{\tiny Periodic}} [D\phi]\exp{\left(\int_0^\beta\,d\tau\int d^3x\,{\cal L}\right)}\,,
\end{equation}
where ${\cal L}$ is the Lagrangian density for the theory under consideration, $N$ is a constant and `periodic' means that the functional integral is to be evaluated over field configurations satisfying:
\begin{equation}\label{PC}
\phi(x,y,z,0)=\phi(x,y,z,\beta)\,, 
\end{equation}
where $\beta=T^{-1}$, the reciprocal of the temperature, is the periodic length in the Euclidean time axis. The Helmholtz free energy $F(\beta)$ is related to the partition function ${\cal Z}(\beta)$ through the relation $F(\beta)=-\beta^{-1}\log{{\cal Z}(\beta)}$. Other than the periodic conditions given by (\ref{PC}), we must also consider boundary conditions which are determined by the geometry and the nature of the physical system under study. An example is the configuration mentioned  above. Choosing Cartesian axes such that the axis $OZ$ is perpendicular to both plates with the perfectly conducting plate at $z=0$ and the infinitely permeable one at $z=d$, the boundary conditions on the vacuum oscillations of the electromagnetic field are the following: the tangential components of the electric field as well as the normal component of magnetic field must vanish at $z=0$, while the tangential components of the magnetic field must vanish at $z=d$. For the plate geometry tha!
t we are considering, the electromagnetic field can be mimicked by a scalar massless field $\phi$. The boundary conditions stated above can be translated into:

\begin{equation}\label{MC}
\phi(\tau, x,y,z=0)=0\,;\;\;\;\;\;{\partial\phi(\tau,x,y,z=d)\over\partial z}=0\,,
\end{equation}
where $\tau$ is the Euclidean time. The insertion at the end of the calculation of a factor $2$ will take into account the two possible transverse polarizations of the electromagnetic field. Thus we write $\log{{\cal Z}(\beta)}$ as:
\begin{equation}\label{PFU}
\log{{\cal Z}(\beta)}=\left(-{1\over 2}\right)\log\det\left(-\partial_{\mbox{\tiny E}}|{\cal F}_d\right)\,,
\end{equation}
where $\partial_{\mbox{\tiny E}}=\partial^2/\partial\tau^2+\nabla^2$, and the symbol ${\cal F}_d$ stands for the set of functions which satisfy conditions (\ref{PC}) and (\ref{MC}). The generalized zeta function method  basically consists of the following three steps: (i) first, we compute the eigenvalues of the operator $-\partial_{\mbox{\tiny E}}$ subject to the appropriate boundary conditions and write $\zeta (s;-\partial_{\mbox{\tiny E}})=\mbox{Tr}\,(-\partial_{\mbox{\tiny E}})^{-s}$; (ii) second, we perform an analytical continuation of $\zeta(s;-\partial_{\mbox{\tiny E}})$ to a meromorphic function on the whole complex $s$-plane; (iii) finally, we compute $\det{(-\partial_{\mbox{\tiny E}}|{\cal F}_d)}=\exp{\left(-{\partial\zeta (s=0;-\partial_{\mbox{\tiny E}})\over\partial s}\right)}$. Combining equation (\ref{PFU}) with the definition of free energy we obtain:
\begin{equation}
F(\beta)=-\beta^{-1}{\partial\zeta (s=0;-\partial_{\mbox{\tiny E}})\over\partial s}\,.
\end{equation}
The eigenvalues of $-\partial_{\mbox{\tiny E}}$ whose eigenfunctions satisfy (\ref{PC}) and (\ref{MC}) are:
\begin{equation}
\left\{k_x^2+k_y^2+\left(n+{1\over 2}\right)^2{\pi^2\over d^2}+{4\pi^2m^2\over \beta^2},\right\},
\end{equation}
where $k_x,k_y \in\Re$, $ n\in \left\{0,1,2,3,...\right\}$ and $m\in\left\{ 0,\pm 1, \pm 2,...\right\}$. The generalized zeta function then reads:
\begin{eqnarray}
\zeta (s,-\partial_{\mbox{\tiny E}})& = & L^2\sum_{m=-\infty}^\infty\sum_{n=0}^\infty\int{dk_xdk_y\over (2\pi)^2}\times \nonumber\\
&\times&\left[k_x^2+k_y^2+(2n+1)^2{\pi^2\over 4d^2}+{4\pi^2m^2\over\beta^2}\right]^{-s}\;.
\end{eqnarray}
where $L^2$ is the area of the plates. After rearranging terms in the summations, changing to polar coordinates and integrating the angular part out, we can rewrite this last equation as:
\begin{eqnarray}
& \zeta & (s,-\partial_{\mbox{\tiny E}}) = {L^2\over 2\pi}\left\{\sum_{n=1,3,5,...}\int_0^\infty dk\,k\left[k^2+{n^2\pi^2\over 4d^2}\right]^{-s}\right. \nonumber \\
& + & \left. 2\sum_{n=1}^{\infty\;\;\prime}\sum_{m=0}^\infty dk\,k\left[k^2+{n^2\pi^2\over 4d^2}+{4\pi^2m^2\over\beta^2}\right]^{-s}\right\}\;,
\end{eqnarray}
where $k^2=k_x^2+k_y^2$ and the prime on the summation symbol serve to remind us that the integer $n$ assumes odd values only. Using the following representation for the Euler beta function, {\it c.f.} formula {\bf 3.251}.2 \cite{Grad}:
\begin{eqnarray}
\int_0^\infty dx\,x^{\mu-1}\left(x^2+a^2\right)^{\nu-1} &=& {B\over2}\left({\mu\over2},1-\nu-{\mu\over2}\right)\nonumber \\
& \times & a^{\mu+2\nu-2}\,,
\end{eqnarray}
where
\begin{equation}
B(x,y)={\Gamma(x)\Gamma(y)\over\Gamma(x+y)}
\end{equation}
which holds for $\Re\,\left(\nu+{\mu\over 2}\right)< 1$ and $\Re\,\mu>0$, we obtain:

\begin{eqnarray}
\zeta (s,-\partial_{\mbox{\tiny E}}) & = & {L^2\over 4\pi}{\Gamma (s-1)\over \Gamma (s)}\left[\left({\pi\over 2d}\right)^{2-2s}\sum_{n=}^{\infty\;\;\prime} n^{2-2s}  \right. \nonumber \\
 & + & \left. 2\pi^{2-2s}\sum_{m=1}^\infty\sum_{n=1}^{\infty\;\;\prime}\left[{n^2\over 4d^2}+{4m^2\over\beta^2}\right]\right]^{1-s}
\end{eqnarray}
In order to connect the simple sum on the r.h.s. of the above equation to the Riemann zeta function $\zeta_R$ we write:

\begin{equation}
\sum_{n=1}^{\infty\;\;\prime} n^{2-2s}=(1-2^{2-2s})\zeta_R(2s-2).
\end{equation}
On the other hand, the double sum can be expressed in terms of Epstein functions which for any positive integer $N$ and $\Re\,z$ large enough are defined by \cite{Epstein,Kirsten}:

\begin{eqnarray}
& E &_N^{M^2}(z;a_1,a_2,...,a_N):=\sum_{n_1=1}^{\infty}
\sum_{n_2=1}^{\infty}...\sum_{n_N=1}^{\infty}\nonumber\\
&&{1\over
(a_1n_1^2+a_2n_2^2+...+a_Nn_N^2+M^2)^z},
\end{eqnarray} 
where $a_1,...a_N$ and $M^2$$>0$ and writing:
\begin{eqnarray}
& & \sum_{m=1}^\infty \sum _{n=1}^{\infty\;\;\prime}\left[{n^2\over 4d^2}+{4m^2\over\beta^2}\right]^{1-s}=\sum_{m,n=1}^\infty\left[{n^2\over 4d^2}+{4m^2\over\beta^2}\right]^{1-s}\nonumber \\ 
& &\;\;\;\;\;\;\;\;\;\;\;\;\;\;\;\;\;\;\;\;\;\;\;\;\;\;\;\;\;\;\;\;\; -\sum_{m, n=1}^\infty\left[{(2n)^2\over 4d^2}+{4m^2\over\beta^2}\right]^{1-s},
\end{eqnarray}
we can write:
\begin{eqnarray}
& &\zeta (s,-\partial_{\mbox{\tiny E}}) = {L^2\over 4\pi}{\Gamma (s-1)\over \Gamma (s)\pi^{2-2s}} \nonumber \\ & &\;\;\;\;\;\;\;\;\;\;\;\;\;\; \times\left[\left({1\over 2d}\right)^{2-2s}\left(1-2^{2-2s}\right)\zeta_R (2s-2)\right. \nonumber \\
& + & \left. 2E_2\left(s-1;{1\over 4d^2}, {4\over\beta^2}\right)-2E_2\left(s-1;{1\over d^2}, {4\over\beta^2}\right)\right].
\end{eqnarray}
The Epstein functions can be analytically continued to a meromorphic function in the complex plane, (see for example \cite{Kirsten}). For $N=2$ and $M^2=0$ the analytic continuation is given by:
\begin{eqnarray}
& & E_2(z;a_1,a_2) = 
-{a_1^{-z}\over 2}\zeta_R (2z)+{1\over 2}\sqrt{\pi\over a_2}
{\Gamma(z-{1\over 2}) \over \Gamma(z)} \nonumber \\
& &\;\;\;\times\; E_1(z-{1\over 2};a_1)
+ {2\pi^z \over \Gamma(z) a_2^{{z\over 2}+{1\over 4}}}\nonumber \\
& &\;\;\;\times\sum_{n,m=1}^\infty 
{m^{z-(1/2)}\over (a_1 n^2)^{(z-(1/2))/2}}
K_{{1\over 2}-z}\left({2\pi m\over \sqrt{a_2}}\sqrt{a_1n^2}\right).
\end{eqnarray}
Here $K_\nu (z)$ is a Macdonald's function\footnote{We use the terminology employed by N. N. Lebedev in {\it Special Functions and Their Applications}, Dover Publications, New York, 1972. The function $K_\nu (z)$ is also known as modified Bessel function of the third kind and Bessel function of imaginary argument.} defined on the complex $z$-plane cut along the negative real axis, $[-\infty, 0]$. Performing the appropriate substitutions for $z$, $a_1$ and $a_2$
and taking advantage of the useful fact that the derivative of the function $G(s)/\Gamma (s)$ at $s=0$ is simply $G(0)$ for a well-behaved $G(s)$ we obtain:

\begin{eqnarray}
&\zeta&^\prime(s,-\partial_{\mbox{\tiny E}}) = -{7\over 8}\times{\pi^2\beta L^2\over 720 d^3}+ {L^2\sqrt{2}\over\sqrt{\beta}} \sum _{n,m=1}^\infty\left({md\over n}\right)^{-{3\over2}}\nonumber \\ 
&\times & \left[2^{-{3\over 2}}K_{3/2}\left({\beta\pi nm\over 2d}\right)-K_{3/2}\left({2\beta\pi nm\over 2d}\right)\right]\,.
\end{eqnarray}

It follows that the Helmholtz free energy per unit area for this peculiar arrangement is given by:
\begin{eqnarray}\label{FR}
{F\over L^2}&=&{7\over 8}\times{\pi^2\over 720 d^3} -{\sqrt{2}\over\beta^{{3\over 2}}}\sum_{n,m=1}^\infty\left({md\over n}\right)^{-{3\over2}}\nonumber \\ &\times &\left[2^{-{3\over 2}}K_{3/2}\left({\beta\pi nm\over 2d}\right)-K_{3/2}\left({2\beta\pi nm\over 2d}\right)\right]\,,
\end{eqnarray}
where we have already accounted for the two possible polarization states. The first term in (\ref{FR}) represents the regularized repulsive Casimir energy at zero temperature found by Boyer \cite{Boyer}. Notice that this term is $-7/8$ times the result obtained for the Casimir effect with Dirichlet boundary conditions at $T=0$. The second term in (\ref{FR}) is the contribution to the free energy due to thermal effects and we can recast it into a more manageable form as is shown next.

The Macdonald's functions $K_\nu (z)$ of half-integral order are given by ({\it c.f.} formula {\bf 8.468} in \cite{Grad}):
\begin{equation}\label{BeH}
K_{n+{1\over 2}}(z)=\sqrt{{\pi\over 2z}}e^{-z}\sum_{k=0}^n{(n+k)!\over k!(n-k)!(2z)^k}\,.
\end{equation}
Hence, defining the dimensionless variable $\xi$ by $\xi:=d/\pi\beta=Td/\pi$ and making use of (\ref{BeH}), we can recast (\ref{FR}) into the form:
\begin{equation}\label{FREE}
{F(\beta)\over L^2}={7\over 8}\times{\pi^2\over 720d^3}-{1\over \pi\beta^3}f(\xi)\,,
\end{equation}
where $f(\xi)$ is a dimensionless function defined by the double sum:
\begin{eqnarray}\label{DS}
f(\xi)& := &\sum_{n,m=1}^\infty\left[\left({1\over m^3}+{n\over 2\xi m^2}\right)e^{-nm/2\xi}\right. \nonumber \\
& - & \left.\left({1\over m^3}+{n\over \xi m^2}\right)e^{-(nm/\xi)}\right]\,.
\end{eqnarray}
The sum over $n$ can be readly evaluated and after some manipulations we end up with:
\begin{eqnarray}\label{FR2}
f(\xi)={1\over 4\xi}\sum_{n=1}^\infty{ \left[{2\xi\over n}+\coth{\left({n\over 2\xi}\right)}\right]\over n^2\sinh{\left({n\over 2\xi}\right)} }\,.
\end{eqnarray}
Equation (\ref{FR2}) summarizes all thermodynamical information concerning the bosonic  excitations confined between the plates.
From (\ref{FR2}) we can easily obtain the low temperature regime of the free energy. It suffices to set $\coth{\left({n\over 2\xi}\right)}\approx 1$, $\sinh{\left({n\over 2\xi}\right)}\approx \exp{{\left({n\over 2\xi}\right)}}$ and keep the term corresponding to $n=1$:
\begin{equation}
f(\xi\ll 1)\approx {1\over 2}\left(1+{1\over 2\xi}\right)\exp{(-{1\over 2\xi})}\,.
\end{equation}
This yields the low temperature limit
\begin{equation}
{F(T)\over L^2}={7\over 8}\times{\pi^2\over 720 d^3}-{1\over 2}\left({T^3\over\pi}+{T^2\over 2d}\right) e^{-\pi/2Td}.
\end{equation}
The very high temperature limit is obtained by setting $\coth{\left({n\over 2\xi}\right)}\approx {2\xi\over n}$ and $\sinh{\left({n\over 2\xi}\right)}\approx {n\over 2\xi}$ and evaluating the sum. The result is:
\begin{equation}
f(\xi\gg 1)\approx {1\over 45}\pi^4\xi .
\end{equation}
This will lead to the Stefan-Boltzmann term corresponding to a slice of vaccum of volume $L^2d$. More accurate  results at high temperature demand that we transform the slowly convergent sum over $m$ in (\ref{DS}) into a more rapidly convergent one. This can be accomplished with the help of Poisson summation formula as we shall see next.
\section{The pressure} 
Let us go back to the scaled free energy $f(\xi)$ defined by the double sum (\ref{DS}). Each term in (\ref{DS}) can be written in the form
\begin{eqnarray}
& & \sum_{n,m=1}^\infty\left({a\over m^3}+{bn\over m^2}\right)e^{-nm/c} = 
\nonumber \\ \;\;\;\;\;\;\; & - & a\beta^2\sum_{n=1}^\infty 
\int_{n\kappa}^\infty d\omega\omega\ln\left({1-e^{-\beta\omega}}\right)\,
\end{eqnarray}
where $a$, $b$ and $c$ are constant satisfying the condition $a=bc$, and $\kappa=1/\beta c$. The scaled free energy can be recast into the form
\begin{eqnarray}\label{HS}
f(\xi) & = & -\beta^2\sum_{n=1}^\infty \int_{n\kappa_1}^\infty d\omega\;\omega\ln\left({1-e^{-\beta\omega}}\right) \nonumber \\
&+&\beta^2\sum_{n=1}^\infty
\int_{n\kappa_2}^\infty d\omega\;\omega\ln\left({1-e^{-\beta\omega}}\right)\,
\end{eqnarray}
where $\kappa_1=\pi/2d$ and $\kappa_2=\pi/d$. The first term (\ref{HS})   corresponds to the thermal correction for two infinite parallel perfectly conducting ($\epsilon\to\infty$) plates separated by a distance $2d$. The second one corresponds to the same setup but with the plates separated by a distance $d$. In this way we can see that the setup we are considering here is the difference between the two setups described above. If we apply this reasoning to the zero temperature term, see (\ref{FREE}), we reproduce the factor $7/8$ with the correct algebraic sign. The net pressure on the plates is given by minus the derivative of the free energy per unit area with respect to the distance $d$ between the plates and like the free energy it splits into the zero temperature contribution and the thermal corrections, that is
\begin{equation}
{\cal P}_{\mbox{\tiny net}}= {7\over 8}{\pi^2\over 240 d^4}+{1\over\pi^2\beta^4}{df(\xi)\over d\xi},
\end{equation}
The thermal contribution reads:
\begin{eqnarray}\label{TC}
{\cal P}_{\mbox{\tiny thermal}} &=& -{1\over\pi^2\beta^4\xi^3}\left[{1\over 4}\sum_{n=1}^\infty n^2\ln{\left(1-e^{-n/2\xi}\right)}\right. \nonumber \\
&-&\left. \sum_{n=1}^\infty n^2\ln{\left(1-e^{-n/\xi}\right)}\right].
\end{eqnarray}
Now we are ready to make use of one of the several versions of Poisson summation formula \cite{Poisson}. The particular version suitable for our purposes reads:
\begin{equation}\label{PoissonEven}
\sum_{n=1}^\infty G(n)=-{G(0)\over 2}+\sum_{l=-\infty}^\infty\int_0^\infty dx\,G(x)\,\cos{(2\pi lx)},
\end{equation}
If we use (\ref{PoissonEven}) in (\ref{TC}) and add the result to the zero temperature contribution we obtain
\begin{eqnarray}\label{NETPRESSURE}
{\cal P}_{\mbox{\tiny net}} & = & {\pi^2\over 45\beta^4}-{1\over 32\pi^4\beta^4}{\partial^2\over\partial\xi^2}{1\over\xi}\sum_{m=1}^\infty{\coth{(4\pi^2 m \xi)}\over m^3} \nonumber \\
&+& {1\over 8\pi^4\beta^4}{\partial^2\over\partial\xi^2}{1\over\xi}\sum_{m=1}^\infty{\coth{(2\pi^2 m \xi)}\over m^3}.
\end{eqnarray}
This result holds for all temperatures. Notice that the zero temperature pressure is apparently missing in our final result. This happens because upon the application of Poisson summation formula we obtain, besides the Stefan-Boltzmann term and the two sums, a term with a negative sign which exactly cancels out the repulsive zero temperature contribution. A similar cancellation occurs also in the case of the Casimir effect for confined massless fermions at finite temperature  \cite{GR} and in the high temperature limit of the standard electromagnetic Casimir effect as shown, for instance, in Plunien {\it et al} \cite{MosteTrunov}. Nevertheless, it is a straightforward matter to show that if we take the zero temperature limit of (\ref{NETPRESSURE}) we  recover the zero temperature term.

The high temperature limit is also easily obtained from (\ref{NETPRESSURE}). Approximating conveniently the hyperbolic cotangents in the sums and evaluating the second partial derivatives and keeping the leading correction terms only we obtain:
\begin{eqnarray}
{\cal P}_{\mbox{\tiny net}}& \approx & {\pi^2T^4\over 45}+{3\zeta (3)T\over 16d^3} \nonumber \\
&+&{1\over 2}{T\over \pi d^3}e^{-4\pi Td} \left(1+4\pi Td+8\pi^2 T^2d^2\right).
\end{eqnarray}
 A simple integration of (\ref{NETPRESSURE}) yields another possible representation for the Helmholtz free energy of this setup:
\begin{eqnarray}\label{HFE}
{F(\beta)\over L^2} & = & -{\pi^2d\over 45\beta^4}+{1\over 32\pi^3\beta^3}{\partial\over\partial\xi}{1\over\xi}\sum_{m=1}^\infty{\coth{(4\pi^2 m \xi)}\over m^3} \nonumber \\
&-& {1\over 8\pi^3\beta^3}{\partial\over\partial\xi}{1\over\xi}\sum_{m=1}^\infty{\coth{(2\pi^2 m \xi)}\over m^3},
\end{eqnarray}
where the integration constant is determined by demanding that in the very high temperature limit the only surviving term in (\ref{HFE}) must be the Stefan-Boltzmann term. Notice that we can also determine this integration constant analyzing the zero temperature limit of (\ref{HFE}). The high temperature limit of (\ref{HFE}) is given by
\begin{eqnarray}
{F(T)\over L^2} & = & -{\pi^2T^4d\over 45}+{3\over 32}{\zeta (3)T\over d^2}\nonumber \\
& + & \left({T\over 4\pi d^2}+{T^2\over \pi d}\right)e^{-4\pi Td}
\end{eqnarray}
Apart from the all-important signs and numerical factors,
these results compare to those obtained in this limit for the attractive case, see for example Plunien {\it et al} \cite{MosteTrunov}.
\section{Conclusions}
In this paper we have shown how neatly the generalized zeta function regularization method applies to the repulsive electromagnetic Casimir effect at finite temperature for the simple geometry of infinite parallel plates with certain special physical properties. As a follow up of the application of this method we have obtained expressions for the Helmholtz free energy and the force per unit area acting on any one of the two plates which comprise this peculiar system. We also obtained the low and high temperature limits of those two quantities.  It almost goes without saying that the standard Casimir effect at zero and finite temperature can also be treated in the same way. 

It is also worth noticing that our high temperature limit result is compatible with ideas of dimensional reduction that occurs for $T$ $\to\infty$. Remark that in this limit, if we omit the Stefan-Boltzmann term,  the dominant term in the pressure is proportional to $1/d^3$, a result reminiscent of $2+1$ dimensions, in contrast to the $1/d^4$ term, typical of $3+1$ dimensions. 

The interesting similarities and differences of the Casimir effect associated with a massless bosonic field, which arise when we compare the consequences of imposing Dirichlet boundary conditions in one case and mixed ones in the other, indicate that a similar investigation in other theories, such as the massive scalar field at zero as well as at finite temperature might be rewarding. This investigation is being carried out and results will be published elsewhere.

\acknowledgments
\rm The authors are indebt to our colleagues Carlos Farina and Ashok Das for carefully reading the manuscript and enlightning discussions. One of us (A. Ten\'orio) wishes to acknowledge the financial support of CNPq, the Brazilian research agency.
\end{document}